\documentclass[journal]{IEEEtran}
\usepackage{amsmath}
\usepackage{graphicx}
\usepackage{url}
\usepackage{algorithm}
\usepackage{algorithmic}

\begin{document}

\title{Deep-Q Learning with Hybrid Quantum Neural Network on Solving Maze Problems}

\author{*Hao-Yuan Chen, Yen-Jui Chang, Shih-Wei Liao, Ching-Ray Chang
\\ Department of Computer Science, University of London, London WC1E 7HU, United Kingdom
\\ Department of Physics, National Taiwan University, Taipei 106216, Taiwan 
\\ Department of Computer Science and Information Engineering, National Taiwan University, Taipei 106216, Taiwan 
\\ Quantum Information Center, Chung Yuan Christian University, Taoyuan 320314, Taiwan
\\ *E-mail: (hc118@student.london.ac.uk)
\thanks{Manuscript Revised at November 2023}}

\maketitle

\begin{abstract}
Quantum computing holds great potential for advancing the limitations of machine learning algorithms to handle higher dimensions of data and reduce overall training parameters in deep learning (DL) models. This study uses a trainable variational quantum circuit (VQC) on a gate-based quantum computing model to investigate the potential for quantum benefit in a model-free reinforcement learning problem. Through a comprehensive investigation and evaluation of the current model and capabilities of quantum computers, we designed and trained a novel hybrid quantum neural network based on the latest Qiskit and PyTorch framework. We compared its performance with a full-classical CNN with and without an incorporated VQC. Our research provides insights into the potential of deep quantum learning to solve a maze problem and, potentially,  other reinforcement learning problems. We conclude that reinforcement learning problems can be practical with reasonable training epochs. Moreover, a comparative study of full-classical and hybrid quantum neural networks is discussed to understand these two approaches' performance, advantages, and disadvantages to deep-Q learning problems, especially on larger-scale maze problems larger than 4x4.
\end{abstract}

\begin{IEEEkeywords}
Quantum Machine Learning, Hybrid Quantum Neural Networks, Deep Reinforcement Learning, Quantum Deep-Q learning
\end{IEEEkeywords}

\IEEEpeerreviewmaketitle

\section{Introduction}
\IEEEPARstart{Q}uantum machine learning is a novel and highly immature field of study. Its promising implication for combining two powerful areas in computer science and physics, quantum machine learning, has made an interesting topic for researchers to investigate to resolve various challenging problems. \cite{biamonte2017} However, considering the limitations of Noisy Intermediate-Scale Quantum (NISQ) devices and machine learning algorithm designs, some experimental models haven’t demonstrated strong effectiveness and eventually surpass the present performance of a full classical model. Therefore, this research would like to investigate the potential and effectiveness of incorporating a variational quantum circuit (VQC) with a deep neural network to solve a reinforcement learning problem. Ultimately, this research would like to provide some potential improvements that could be made within this hybrid deep learning model. Also, the study is based on the IBM Quantum Systems (IBM-Q) ecosystem to enable rapid testing and evaluation. The quantum computation experiments were tested and trained on the full-classical hardware with IBM Qiskit’s Aer simulator in the ideal, noise-free environment. 

\subsection{Quantum Reinforcement Learning}
Based on the current quantum machine learning study, there are two approaches to the reinforcement learning problem; the first is to utilize a quantum walk algorithm and variational quantum circuit (VQC) to encode classical agents and environments into a quantum state of information. The second approach, however, incorporates the latest VQC to explore potential speed-up or model size reduction for the present deep learning models with quantum deep learning models.

\subsubsection{Full-Quantum Reinforcement Learning}
Quantum reinforcement learning using a full-quantum solution has been discussed before in 2022. \cite{dalla2022quantum} The research uses a discrete-time quantum walk to map the state and action space into quantum states, resulting in a quantum maze problem. However, considering the constraints of NISQ devices, there is a limited practical application for this type of solution in the near term. Although the research contains limitations to the scalability, it has demonstrated substantial implications for the present full-quantum reinforcement learning algorithm that can provide speed-up for training under a small problem size, six by six maze size.

\subsubsection{Trainable VQC Model}
Another approach to this problem is to build a deep neural network with a variational quantum circuit (VQC). \cite{chen2020} The research, different from this paper, introduces a VQC as a trainable model without any conventional tensor layers for the frozen-lake environment, which is logically equivalent to the maze problem set used in this research. The research has demonstrated that a well-designed VQC is a trainable model for a reinforcement learning problem. Moreover, the model has shown a strong parameter reduction capability from around \(O(n^3)\) and \(O(n^2)\) to \(O(n)\).

Considering the limitations of NISQ devices, near-term engineering applications on the first approach will be regarded as impractical. However, the performance of a trainable VQC would massively depend on the circuit design. Therefore, the research proposed a novel architecture from the previous study \cite{dalla2022quantum} with the latest IBM Qiskit Quantum Primitive to explore other design variants. Therefore, this research suggested a novel integration of quantum gate primitive with the Qiskit framework and conventional tensor layers as the third scheme to solve this problem class as a much generalized deep learning algorithm.

\section{Reinforcement Learning}
Reinforcement learning is the third paradigm of machine learning, which involves an agent learning to make a sequence of decisions in an environment to maximize a cumulative reward signal. The agent interacts with the environment by taking action and receiving rewards. The goal is to learn a policy that maps states to actions to maximize the expected cumulative reward over time.

The fundamental mathematical reinforcement learning model can be formalized as a Markov decision-making process (MDP): At each time step t, the agent observes a state \(s_t\) of the environment. The agent selects an action \(a_t\) from a set of possible actions based on the observed state. The action is executed in the environment, and the agent receives a reward \(r_t\) and transitions to a new state \(s_{t+1}\).

Using a learning algorithm, the agent updates its policy based on the observed state, selected action, received a reward, and new state. A performance metric guides the reward signal learning algorithm, which assigns a scalar value to each state-action pair based on its desirability. The agent aims to learn a policy that maximizes the expected cumulative reward over time, defined as the sum of the rewards received over a finite or infinite horizon, discounted by a factor gamma.

The mathematical model of RL can be represented as a Markov decision process (MDP), which is a tuple
\begin{equation}
    (S, A, P, R, \gamma)
\end{equation}
Where: S is a set of states where the environment can be. A is a set of actions that the agent can take. Given the current state and action, p is a probability distribution over the next state. R is a reward function that assigns a scalar reward to each state-action pair. Gamma is a discount factor that determines the importance of future rewards relative to immediate rewards.
The goal of RL is to find a policy pi(s) that maps each state to a probability distribution over actions, such that the expected cumulative reward over time is maximized:
\begin{equation}
J(pi) = E[R_0 + \gamma_1 + \gamma^2R_2 + ... | s_0, pi],    
\end{equation}
where\(R_t\) is the reward received at time step t, and\(s_0\)is the initial state of the computation.

RL algorithms can be categorized into model-based and model-free approaches. Model-based approaches learn a model of the environment, including the transition probabilities and reward function, and use this model to compute an optimal policy. Model-free methods, such as Q-learning and SARSA, directly estimate the value of the state-action pairs and use this estimate to update the policy. In this research, the researcher proposed integrating a novel parameterized quantum circuit and hybrid neural network to better approximate the relation of state-action based on the model-free approach framework.

\subsection{Q-Learning}
Q-learning is a model-free reinforcement learning algorithm that learns an optimal policy for an agent in an environment with discrete states and actions. It is called "Q-learning" because it learns an estimate of the Q-value function, which is the expected cumulative reward for taking a step in a given state and following the optimal policy afterward.

The Q-value function is estimated using a table or function approximator. The Q-learning algorithm starts with an initial Q-value function, which is gradually updated as the agent interacts with the environment. The agent observes the current state of the environment, selects an action using an exploration-exploitation strategy (such as epsilon-greedy), and receives a reward from the environment. Based on this reward and the resulting state, the Q-value function is updated using the Bellman equation:

\begin{align}
Q(s_t, a_t)\leftarrow \nonumber \\ Q(s_t, a_t) + \alpha \left[r_{t+1} + \gamma \max_{a} Q(s_{t+1}, a) - Q(s_t, a_t)\right]
\end{align}

\subsection{Deep-Q Learning}
Deep Q-learning is a reinforcement learning algorithm that uses deep neural networks to approximate the Q-function, which measures the expected cumulative reward for taking action in a particular state. The Q-function is learned through an iterative process where the agent takes actions in the environment and receives rewards. The agent uses the experiences gathered from these interactions to update its estimates of the Q-function using a technique called temporal-difference learning.

In deep Q-learning, a deep neural network approximates the Q-function. The neural network takes the state of the environment as input and outputs a vector of Q-values for each possible action. The agent then chooses the action with the highest Q-value and executes it in the environment.

The neural network training minimizes the difference between the predicted and target Q-values obtained using the Bellman equation. The Bellman equation expresses the expected cumulative reward for taking action in a particular state as the sum of the immediate reward and the discounted expected cumulative reward from the next state. The loss function used to train the model adopts the mean square error function to calculate the loss of the deep neural network's current parameter configuration.
\begin{equation}
L = \left(r + \gamma \max_{a'} Q(s', a'; \theta_{target}) - Q(s, a; \theta)\right)^2
\end{equation}

The computational complexity of solving a maze problem using Q-learning and deep Q-learning depends on several factors, such as the size of the maze, the complexity of the environment, the number of actions available to the agent, and the number of episodes or iterations required to converge to an optimal policy.

In Q-learning, the agent learns the optimal policy by iteratively updating a Q-value table, which stores the expected reward for each state-action pair. The algorithm requires a significant amount of memory to store the Q-value table, which grows linearly with the size of the maze and the number of possible actions. The time complexity of Q-learning is proportional to the number of iterations required to converge to an optimal policy, which can be significant for complex environments.

On the other hand, deep Q-learning uses a deep neural network to approximate the Q-value function, reducing the memory requirement and allowing for more efficient learning. However, the time complexity of deep Q-learning is proportional to the number of training iterations required to train the neural network, which can be considerable for large-scale environments. In addition, the computational complexity increases with the neural network architecture's complexity and the input data's size.

Solving a maze problem using Q-learning or deep Q-learning can be computationally expensive, especially for large-scale environments. However, advances in hardware and software technology have made it possible to implement these algorithms efficiently and effectively, making them useful for various applications in reinforcement learning and robotics.

As parameterized quantum circuits (PQCs) and deep neural networks (DNNs) share some similarities in their architecture and learning processes, which makes them helpful in solving similar types of problems, the research integrates both models using Qiskit's PyTorch Connector interface to facilitate a hybrid deep neural network. The study would like to investigate the potential complexity reduction for using PQC-DNN architecture to reduce the energy and computational resources needed for the training. 

Therefore, this research proposed a novel architecture to approach a deep-Q learning problem using a hybrid deep neural network in combination with a deep neural network and a parameterized quantum circuit. The development process includes the following components: environment and training agent.

\subsection{Environment}
This is the system or simulation that the agent will interact with. The environment provides the agent with observations and receives actions from the agent. The environment can be anything from a simple game to a complex physical system.

\subsection{Agent}
Firstly, state preparation is needed to encode the current state of the environment so that it can be fed into the agent's neural network. This can be done using various techniques, such as raw pixels or more abstract features. Next, the agent uses its neural network to select an action based on the current state. The neural network takes the state as input and outputs a set of Q-values representing the expected future rewards for each possible action. The agent selects the action with the highest Q-value (or a random action with some probability to encourage exploration). After that, the agent chooses an action; the environment provides a reward signal based on the new state and the selected action. This reward is used to update the neural network's Q-values. In addition, the agent could facilitate a replay-memory mechanism to store its experiences (state, action, reward, next state) in a replay buffer. The replay buffer randomly samples past experiences, which are then used to update the neural network's Q-values. This helps to prevent the agent from over-fitting to recent experiences.

The most critical part of the agent's training process is the neural network that estimates deep-Q value. The agent's neural network is a deep neural network that inputs the current state and outputs a set of Q-values for each possible action. The network is trained using a combination of supervised learning (to minimize the difference between the predicted Q-values and the actual rewards) and reinforcement learning (to encourage the network to learn from its own experiences).

A complete training loop is needed to allow the agent to interact with the environment. The training loop is an iterative process to search for the optimal Q-value in this research. During the training loop, the agent interacts with the environment, stores its experiences in the replay buffer, samples experiences from the replay buffer to update the neural network, and updates the target network periodically. The training loop continues until the agent reaches a satisfactory level of performance.

\begin{figure}
\centering
\includegraphics[width=0.4\textwidth]{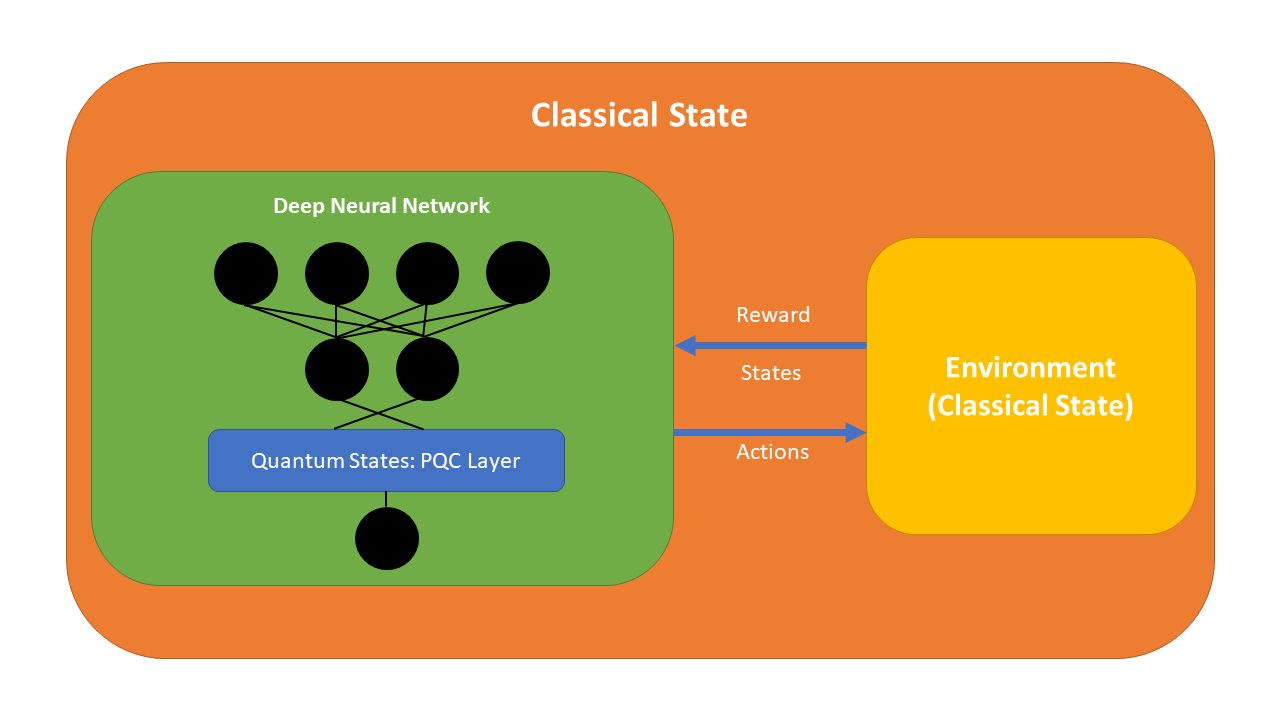}
\caption{Overall research architecture, the environment, and the agent are mostly at classical states of information with determinism using binary encoding. However, the parameterized quantum circuit (PQC) is the quantum state of information that helps the deep neural network estimate the optimal Q-value of deep reinforcement learning problems.}
\end{figure}

\section{Experiment Environment}
Next, the experiment is set up via a classic maze problem. A maze problem refers to finding a path from a starting point to a goal point in a maze or labyrinth. A maze can be represented as a grid of cells, where each cell can be either a wall or a passage. The goal of the maze problem is to find a path from the starting cell to the exit cell while avoiding the walls. The maze problem can be modeled mathematically using a graph with an adjacent matrix. Each node in the graph represents a location in the maze, and each edge represents a possible path between two locations. We can define the graph using an adjacency matrix $A$, where $A_{ij}$ is one if there is a path from node $i$ to node $j$, and 0 otherwise. Figure 2 shows the initial state setting for the agent's policy setting. The hybrid quantum neural network aims to find the optimal policy map.

\begin{figure}
\centering
\includegraphics[width=0.4\textwidth]{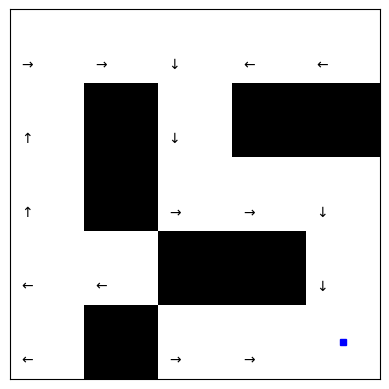}
\caption{The optimal policy defined by the trained hybrid quantum neural network demonstrates the significant effectiveness of the architecture. The arrow in the block means the next direction the agent should take by trained deep-Q network (DQN). The blue point is the exit of the maze.}
\end{figure}

\section{Methods}
Figure 1 depicts the system's overall architecture, which comprises a classical neural network and a trainable variational quantum circuit. This system aims to identify the optimal state-action pairs based on feedback from the environment. The hybrid neural network presented in this study incorporates two primary components: traditional neural layers with multiple tensors and a parameterized quantum circuit integrated into the last layer of the neural network. Hence, designing a hybrid network entails the development of both a PQC and a deep neural model to establish a generalized model for deep-Q learning problems.

\subsection{Hybrid Quantum Neural Network}
Table 1 shows how the hybrid QNN was constructed for the maze size of 4 by 4. The overall model architecture consists of three segments: classical convolutional layers that extract the features of the environment and system feedback into smaller and more digestible information for quantum neural networks that approximate the deep-Q function of the reinforcement learning problems. The third segment is a simple dense layer that learns from the probability distribution of the QNN based on the Sampler primitive to a stable and actionable action space.

\begin{table}[h]
\centering
\begin{tabular}{|l|l|l|}
\hline
\textbf{Layer (type)}    & \textbf{Output Shape}     & \textbf{Param \#} \\ \hline
Conv2d-1         & [-1, 16, 3, 3]   & 80       \\ \hline
ReLU-2           & [-1, 16, 3, 3]   & 0        \\ \hline
Conv2d-3         & [-1, 32, 2, 2]   & 2,080    \\ \hline
ReLU-4           & [-1, 32, 2, 2]   & 0        \\ \hline
Linear-5         & [-1, 2]          & 258      \\ \hline
TorchConnector-6 & [-1, 4]          & 4        \\ \hline
Linear-7         & [-1, 4]          & 20       \\ \hline
\end{tabular}
\vspace{10pt}
\caption{Hybrid Quantum Neural Network Architecture (Maze Size: 4x4)}
\end{table}

\subsection{Parameterized Quantum Circuit}
Next, a parameterized quantum circuit is designed with the IBM Qiskit framework. This code sets up a quantum neural network (QNN) using IBM's latest Sampler-QNN module. The QNN consists of a quantum circuit created using the feature map and Ansatz modules and is defined over two qubits, as shown in Figure 3.

\subsubsection{Quantum Feature Map}
The Pauli-Z evolution circuit, as used in the context of a ZFeatureMap in quantum computing, is a specific type of quantum circuit that employs the Pauli-Z gate to encode classical data into a quantum state. This type of feature map is commonly used in quantum machine learning. A feature map in quantum computing is a method to encode classical data into quantum states. The ZFeatureMap explicitly uses the properties of the Pauli-Z gate to perform this encoding.

\subsubsection{Ansatz}
The study first encodes classical data into quantum states using a quantum feature map, such as a Z-feature map. Then, a Real-Amplitudes circuit is employed as the quantum Ansatz to approximate the Deep-Q function's target distribution. This circuit, commonly used in quantum machine learning, alternates between rotation (Ry gates) and entanglement layers (typically X gates), allowing customizable quantum state preparation. Named "Real Amplitudes" for its generation of states with only real amplitudes. The research configures the circuit without entanglement, resulting in speed-up and efficiency improvement for the quantum simulation.

\subsubsection{Quantum States Measurement}
The Ansatz module, Real-Amplitudes, is a circuit that applies layers of parameterized rotations and entangling gates to the input state. It has two qubits and one repetition of the circuit layer. The Quantum-Circuit (Figure 3) function defines an open quantum circuit of two qubits. Finally, the Sampler QNN module defines a quantum neural network (QNN) that uses the quantum circuit as the model. The feature map and Ansatz parameters specify the circuit's input and weight parameters. The input gradient parameter is set to True, enabling the hybrid gradient backpropagation algorithm for training the QNN. This allows gradients to be backpropagated from the output of the QNN to the input parameters of the feature map module.

\begin{figure}
\centering
\includegraphics[width=0.4\textwidth]{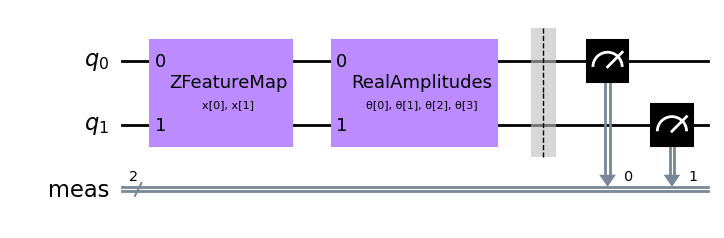}
\caption{A conceptual representation of the variational quantum circuit for this research}
\end{figure}

\subsection{Training Algorithm}
The training algorithm for this hybrid deep neural network is a variant of the Q-learning algorithm, a widely used model-free reinforcement learning algorithm. The algorithm uses a deep neural network as a function approximator to estimate the optimal Q-value function, which maps state-action pairs to their expected future reward. Optimization minimizes the mean squared error between the predicted and target Q-values, computed using a Bellman equation.

The algorithm also employs experience replay, where past transitions are stored in a buffer and sampled randomly during training. This reduces the correlation between consecutive samples and stabilizes the training process. The agent updates the neural network weights using the AdamW optimizer, which adapts the learning rate based on the gradient data with a stabilized approach using weight decay techniques.

The algorithm takes as input the size of the batch, the discount factor gamma, and the device for running the computations. The loss function is the Q-loss, which calculates the mean squared error between the predicted and target Q-values. The \(optimizer.zero_grad()\) method clears the gradients of the optimizer, and the backward() method computes the gradients of the loss function to the neural network parameters. Finally, the optimizer.step() method updates the neural network weights using the calculated gradients.

\begin{algorithm}
\caption{Hybrid QNN Model Training Method}
\begin{algorithmic}
    \STATE Initialize the deep learning model instance 
    \STATE Pick a random state in the environment
    \FOR{$i=0$ to $Number of Epoch$}
        \STATE Zero out the gradients from the previous batch
        \STATE Set the gradients of all the parameters in the optimizer to zero using the method 
        \STATE Sample a batch of data from the agent's replay buffer using the method
        \STATE Calculate the Q-learning loss for the batch of data using the neural network. The discount factor
        \STATE Compute the gradients of the loss to the network parameters using backpropagation. The gradients are then stored in the parameters'
        \STATE Update the network parameters using the gradients computed in the previous step and the optimizer's update rule using the method
    \ENDFOR
\end{algorithmic}
\end{algorithm}

\section{Results}
The research was trained and evaluated on the local machine with NVIDIA's CUDA acceleration framework on the RTX 3080 Ti GPU. In addition to the training hardware, the study applies PyTorch and IBM Qiskit to develop the Hybrid-QNN and CNN. This section presents the training profiles of classical and hybrid neural networks for the maze size ranging from 4x4 to 5x5. Later, the reward history of the classical and quantum neural networks with different problem sizes, ranging from 4x4 to 5x5, are delivered as well to understand the training process of these models.

Moreover, two tables of benchmarking results were presented to evaluate the performance of the classical and quantum deep learning models proposed in this research. The benchmark results were evaluated based on the three metrics: model size, win rates, and training duration required to reach optimal policy states within the 4x4 and 5x5 maze problems.

\subsection{Training Profile}
The model training configuration is an epsilon profile, as shown in Figures 4 and 5. In reinforcement learning, an agent interacts with an environment and learns to take actions that maximize a cumulative reward signal. One common strategy for balancing the exploration of new actions with the exploitation of known good actions is the epsilon-greedy strategy. This strategy involves choosing a random action with probability epsilon and choosing the action with the highest expected reward with probability (1-epsilon).

\begin{figure}
\centering
\includegraphics[width=0.4\textwidth]{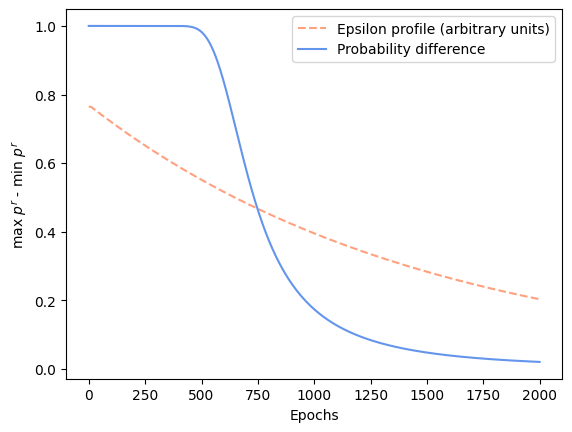}
\caption{Model's Epsilon profile - Maze size: 4x4}
\end{figure}

\begin{figure}
\centering
\includegraphics[width=0.4\textwidth]{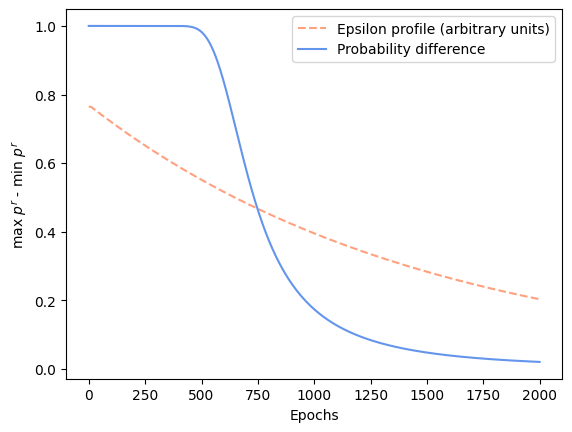}
\caption{Model's Epsilon profile - Maze size: 5x5}
\end{figure}

The value of epsilon is typically set to decrease over time, resulting in an "epsilon profile" that describes how the agent's exploration behavior changes as it gains more experience. Initially, the agent may explore more (i.e., set epsilon to a higher value) to discover new actions that may lead to high rewards. As the agent learns which actions are most likely to lead to high rewards, it may reduce its exploration (i.e., set epsilon to a lower value) to focus more on exploiting known good actions.

\subsection{Training Results}
Table 2 demonstrates the effectiveness of the hybrid quantum neural network introduced in the research from the perspectives of model size, win rates, and total training duration required to train both classical and hybrid QNN agents. The total number of parameters available evaluated the model size. The win rate was calculated by the total epochs and the number of win epochs to evaluate the effectiveness of the training.
\begin{equation}
\text{Win Rate (\%)} = \frac{\text{Win counts}}{\text{Total epochs}} \times 100
\end{equation}

\begin{table}[ht]
\centering
\begin{tabular}{|l|r|r|r|}
\hline
\textbf{Model} & \multicolumn{1}{l|}{\textbf{Model Size}} & \multicolumn{1}{l|}{\textbf{Win Rate}} & \multicolumn{1}{l|}{\textbf{Training Runtime}} \\ \hline
Classical CNN & 6588 & 89.94\% & 88.29 sec \\ \hline
Hybrid QNN & 2442 & 85.19\% & 700.28 sec \\ \hline
\end{tabular}
\hspace{1.5cm}
\caption{Comparison of Model Performances on 4x4 Maze Size}
\end{table}

\begin{table}[ht]
\centering
\begin{tabular}{|l|r|r|r|}
\hline
\textbf{Model} & \multicolumn{1}{l|}{\textbf{Model Size}} & \multicolumn{1}{l|}{\textbf{Win Rate}} & \multicolumn{1}{l|}{\textbf{Training Runtime}} \\ \hline
Classical CNN & 7114 & 94.87\% & 297.86 sec \\ \hline
Hybrid QNN & 4890 & 93.13\% & 1577.37 sec \\ \hline
\end{tabular}
\hspace{1.5cm}
\caption{Comparison of Model Performances on 5x5 Maze Size}
\end{table}

\begin{figure}
\centering
\includegraphics[width=0.4\textwidth]{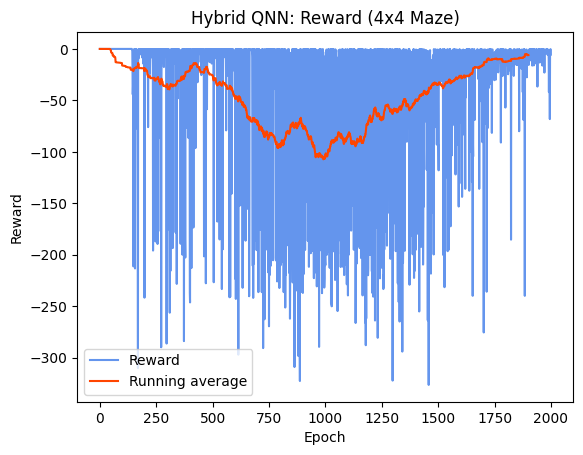}
\caption{Hybrid QNN's Reward History: Maze Size (4x4)}
\end{figure}

\begin{figure}
\centering
\includegraphics[width=0.4\textwidth]{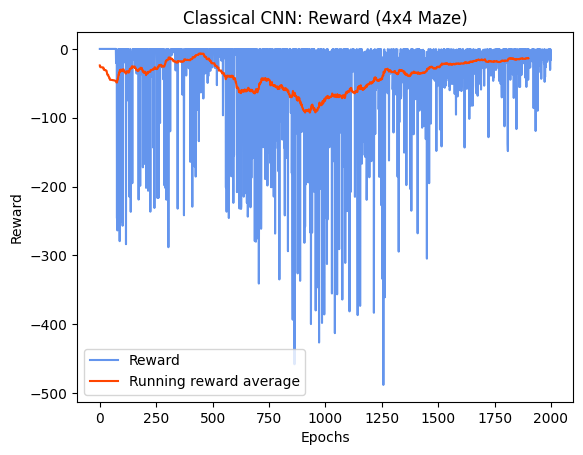}
\caption{Classical CNN's Reward History: Maze Size (4x4)}
\end{figure}

\begin{figure}
\centering
\includegraphics[width=0.4\textwidth]{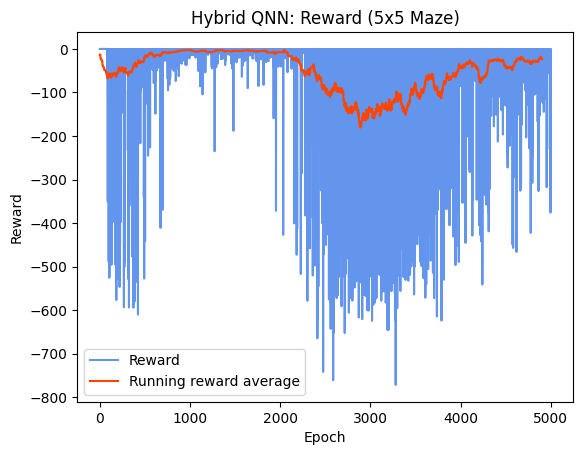}
\caption{Hybrid QNN's Reward History: Maze Size (5x5)}
\end{figure}

\begin{figure}
\centering
\includegraphics[width=0.4\textwidth]{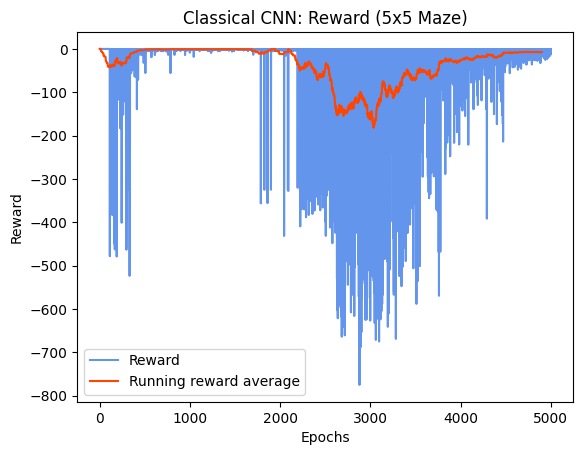}
\caption{Classical CNN's Reward History: Maze Size (5x5)}
\end{figure}

Ultimately, our hybrid deep neural network demonstrates the promising capability and potential of near-term quantum deep learning solutions using a generalized VQC design. The findings suggest that the proposed hybrid network can perform highly in various applications. Using a combination of classical and quantum machine learning techniques, the researchers demonstrated the potential for achieving improved performance in challenging problems. 

These results suggest hybrid deep learning approaches using generalized VQC designs could be vital in developing future quantum machine learning applications. Moreover, as this is a quantum-classical scheme, the solution can be trained on the GPU-accelerated hardware with IBM Aer and NVIDIA's cuQuantum library to explore future applications based on this model architecture.

\section{Discussion}
The research provides the third architecture for approaching quantum reinforcement learning problems using a hybrid neural network. This section compares two other architectures, full-quantum and trainable VQC models.

\subsection{Differences from full-quantum model}
The fundamental difference between this research and the full-quantum model is the state of action and environment. The full-quantum model encodes all the state and action space into a quantum state using a discrete-time quantum walk with a hopping mechanism. The research demonstrates a strong performance and high level of parallelism possible for a full-quantum solution. The degree of complexity reduction is \(O(\sqrt{n})\). However, this solution is most sensitive to the noise within the quantum devices with the lowest scalability potential. 

\subsection{Differences from trainable VQC model}
Therefore, a trainable quantum circuit model is proposed to estimate the Q-value for the Bellman equation. Unlike our research, the VQC's approach doesn't include any classical tensor layers within the model, which increases the difficulty of generalizing the model into larger problem sizes. However, this research has provided a theoretical and empirical foundation for the logical similarity between VQC and DNN.

\subsection{Hybrid Neural Network}
As a result, a hybrid deep neural network incorporating the VQC model is introduced to mitigate the challenges from the previous two related research. With multiple classical neural layers, the model could resolve many generalized problems with various input dimensions. Also, as this neural network is full-trainable and functional on the classical device, a strong implication and potential for near-term applications emerged. However, some technical constraints are still involved within this solution architecture, like quantum-classical gradient descent, accelerated computing devices for this model class, etc. 

\subsection{Outlook}
The challenges of this research are exploring an efficient training method on classical simulators and real quantum hardware. However, in light of current constraints to hardware accessibility and software architecture, training a hybrid model on a real quantum computer is challenging. However, a well-architected training scheme can be designed on the GPU cluster with NVIDIA cuQuantum, IBM Qiskit, and PyTorch framework to reduce the amount of intercommunication between cores and processors, which results in the slow training for this model based on the various iterations of this experiments.

\section{Summary}
This research presents an introduction and technological demonstration of the quantum reinforcement learning algorithm with a hybrid neural network. The neural network comprises the latest circuit architecture with novel IBM Quantum Computing Primitive built-in. The results imply the future trajectory and application of variational quantum circuits (VQC) to accomplish an even more complex environment for deep reinforcement learning agent training. In addition, the research demonstrates how near-term quantum devices could provide potential speed-up or parameter reduction to the current deep learning model. In the end, the study would like to integrate a split-steps quantum walk circuit model with a deep neural network and VQC to load the classical state-space into a quantum state of information in which the current algorithm might provide a further advance in terms of speed and model size reduction for such applications.

\section*{Acknowledgment}
The authors thank Ph.D. candidate Yen-Jui Chang at National Taiwan University for supporting and advising this project. Also, a big thanks to Prof. Shih-Wei Liao and Prof. Ching-Ray Chang for providing professional guidance on research trajectories. We acknowledge support from the National Science and Technology Council, Taiwan, under Grants NSTC 112-2119-M-033-001, for the research project Applications of Quantum Computing in Optimization and Finances.

\section*{Declarations}
\subsection*{Competing interests}
No, I declare that the authors have no competing interests as defined by Springer or other interests that might be perceived to influence the results and discussion reported in this paper.

\subsection*{Authors' contributions}
Mr. Hao-Yuan Chen oversaw the entire research project, taking charge of various critical aspects. He skillfully designed the neural network architecture and conducted rigorous validation experiments. Furthermore, he contributed significantly to crafting most of the paper's contents. Alongside Mr. Hao-Yuan Chen, Prof. Chang and the Ph.D. candidate, Yen-Jui, were valuable collaborators. They actively engaged in stimulating and thought-provoking discussions, enriching the research process. Additionally, their efforts were focused on improving the quantum circuit design, which proved pivotal to the study's success.

\ifCLASSOPTIONcaptionsoff
  \newpage
\fi

\section*{Data Availability}
All data used for this experiment is available on GitHub: https://github.com/MarkCodering/Deep-Reinforcement-Learning-using-quDNN

\end{document}